\documentclass[a4paper,twoside,12pt]{article}
\input{obsjkt.sty}

\newcommand{\Msun}{\ensuremath{~{\rm M}_\odot}}                   
\newcommand{\Rsun}{\ensuremath{~{\rm R}_\odot}}                   
\newcommand{\rhosun}{\ensuremath{~\rho_\odot}}                    
\newcommand{\Teff}{\ensuremath{T_{\rm eff}}}                      
\newcommand{\Grp}{\ensuremath{G_{\rm RP}}}                        
\newcommand{\degr}{\ensuremath{^\circ}}                           
\renewcommand{\kms}{~km~s$^{-1}$}                                 
\newcommand{\as}{\ensuremath{^{\prime\prime}}}                    
\newcommand{\chir}{\ensuremath{\chi_\nu^{\,2}}}                   

\newcommand{\etal}{\textit{et al.}}                               
\newcommand{\corot}{\textit{CoRoT}}
\newcommand{\kepler}{\textit{Kepler}}

\newcommand{\gaia}{\textit{Gaia}}
\newcommand{\targ}{IT~Cas}
\newcommand{\targfull}{IT~Cassiopeiae}

\newcommand{\Msunnom}{\hbox{$\mathcal{M}^{\rm N}_\odot$}}
\newcommand{\Rsunnom}{\hbox{$\mathcal{R}^{\rm N}_\odot$}}
\newcommand{\Lsunnom}{\hbox{$\mathcal{L}^{\rm N}_\odot$}}

\usepackage{rotating}

\begin{document} 

\OBSheader{Rediscussion of eclipsing binaries: \targ}{J.\ Southworth}{2023 June}

\OBStitle{Rediscussion of eclipsing binaries. Paper XIII. \\ The F-type twin system IT Cassiopeiae}

\OBSauth{John Southworth}

\OBSinstone{Astrophysics Group, Keele University, Staffordshire, ST5 5BG, UK}


\OBSabstract{\targ\ is a detached eclipsing binary system containing two F3\,V stars in an orbit of period 3.90~d and eccentricity 0.089. Light curves are available from three sectors of observations from the Transiting Exoplanet Survey Satellite (TESS), and extensive radial velocity measurements have been published by Lacy \etal\ \cite{Lacy+97aj}. We model these data using the {\sc jktebop} code to determine the physical properties of the system. We find masses of $1.324 \pm 0.009$ and $1.322 \pm 0.008$\Msun, and radii of $1.555 \pm 0.004$ and $1.551 \pm 0.005$\Rsun. The two stars are identical to within the uncertainties, and the depths of the primary and secondary eclipses are also indistinguishable. Using the effective temperature of $6740 \pm 105$~K from Lacy \etal\ (for both stars) gives a distance to the system of $505.5 \pm 8.3$~pc, in good agreement with the value of $515.0 \pm 4.4$~pc from the \gaia\ DR3 parallax. The properties of the stars are consistent with theoretical predictions for a solar chemical composition and an age of 2~Gyr. No pulsations are apparent in the TESS photometry.}


\section*{Introduction}

We are currently pursuing a project to determine precise and accurate masses and radii for detached eclipsing binaries (dEBs) for which high-quality spectroscopic orbits and new light curves from the Transiting Exoplanet Survey Satellite (TESS) are available \cite{Me20obs}. The immediate aims of the project are to use the new TESS data to improve the measurements of the properties of dEBs that are already in the Detached Eclipsing Binary Catalogue (DEBCat\footnote{\texttt{https://www.astro.keele.ac.uk/jkt/debcat/}}, ref.~\cite{Me15aspc}), and to add new dEBs to this catalogue. The longer-term aims are to use dEBs to improve our understanding of stellar physics and to help calibrate theoretical evolutionary models \cite{ClaretTorres16aa,Tkachenko+20aa}.

The availability of high-quality light curves from space missions such as \corot\ \cite{Auvergne+09aa}, \kepler\ \cite{Borucki16rpph} and TESS \cite{Ricker+15jatis} has revolutionised many areas of stellar physics \cite{ChaplinMiglio13araa,WinnFabrycky15araa}. Their effect has been keenly felt for binary stars \cite{Lampens21galax,Me21univ,Borkovits22galax}, with light curves of a quality unachievable from the ground now available for thousands of dEBs \cite{IJspeert+21aa,JustesenAlbrecht21apj,Prsa+21apjs} of which many have an extensive observational history.

One of these number is \targfull\ (Table\,\ref{tab:info}), an F-type dEB which has been observed using TESS and for which high-quality radial velocity (RV) measurements have been published. \targ\ was discovered by Fadeeva using photographic plates from Moscow \cite{Parenago39pz,KhaliullinKozyreva89apss}. Photometric analyses have been carried out by several authors since \cite{KhaliullinKozyreva89apss,Florja39spvs,Whitney57aj,Busch75mbhbs,Lacy+95ibvs,HolmgrenWolf96obs}. The system has a small orbital eccentricity and exhibits apsidal motion \cite{KhaliullinKozyreva89apss,HolmgrenWolf96obs,KozyrevaZakharov01astl,Claret+21aa} with a period of $U = 877 \pm 78$~yr \cite{Claret+21aa}.

Lacy \cite{Lacy84ibvs} obtained high-resolution spectra which showed the system to be double-lined, with both components having narrow spectral lines indicative of low rotational velocity. Lacy \etal\ \cite{Lacy+97aj} (hereafter L97) subsequently presented extensive new spectroscopy and photometry from which they determined the physical properties of the system; these include the only mass and radius measurements published so far. The available light curves had poor coverage of the first and last contact points of the eclipses. In this work we use a space-based light curve and the radial velocities (RVs) of L97 to obtain improved measurements of the physical properties of \targ.

Khaliullin \& Kozyreva \cite{KhaliullinKozyreva89apss} detected periodic variability in their light curve of a secondary eclipse, and deduced that the primary component was a $\delta$~Scuti pulsator.
Holmgren \& Wolf \cite{HolmgrenWolf96obs} detected periodic variability in their light curve of a \emph{primary} eclipse, and deduced that the \emph{secondary} component was a $\delta$~Scuti pulsator.
Lacy \etal\ \cite{Lacy+97aj} did not confirm either variation and suggested that they were erroneous. Our own analysis below finds no evidence for short-period variability.
Kozyreva \etal\ \cite{KozyrevaZakharov01astl} found slow variations in the brightness of \targ\ with a timescale of about one month. They found them to occur independent of the choice of comparison star so attributed them to the dEB.


\begin{table}[t]
\caption{\em Basic information on \targ. \label{tab:info}}
\centering
\begin{tabular}{lll}
{\em Property}                            & {\em Value}                 & {\em Reference}                   \\[3pt]
Right ascension (J2000)                   & 23:42:01.38                 & \cite{Gaia21aa}                   \\
Declination (J2000)                       & +51:44:36.8                 & \cite{Gaia21aa}                   \\
\textit{Tycho} designation                & TYC 3650-959-1              & \cite{Hog+00aa}                   \\
\textit{Gaia} DR3 designation             & 1944868020357285504         & \cite{Gaia21aa}                   \\
\textit{Gaia} DR3 parallax                & $1.9419 \pm 0.0165$ mas     & \cite{Gaia21aa}                   \\          
TESS\ Input Catalog designation           & TIC 26801525                & \cite{Stassun+19aj}               \\
$U$ magnitude                             & $11.631 \pm 0.020$          & \cite{Lacy92aj}                   \\
$B$ magnitude                             & $11.640 \pm 0.013$          & \cite{Lacy92aj}                   \\          
$V$ magnitude                             & $11.152 \pm 0.010$          & \cite{Lacy92aj}                   \\          
$J$ magnitude                             & $10.212 \pm 0.020$          & \cite{Cutri+03book}               \\
$H$ magnitude                             & $ 9.957 \pm 0.021$          & \cite{Cutri+03book}               \\
$K_s$ magnitude                           & $ 9.915 \pm 0.016$          & \cite{Cutri+03book}               \\
Spectral type                             & F3\,V + F3\,V               & This work                         \\[3pt]
\end{tabular}
\end{table}


\section*{Observational material}

\begin{figure}[t] \centering \includegraphics[width=\textwidth]{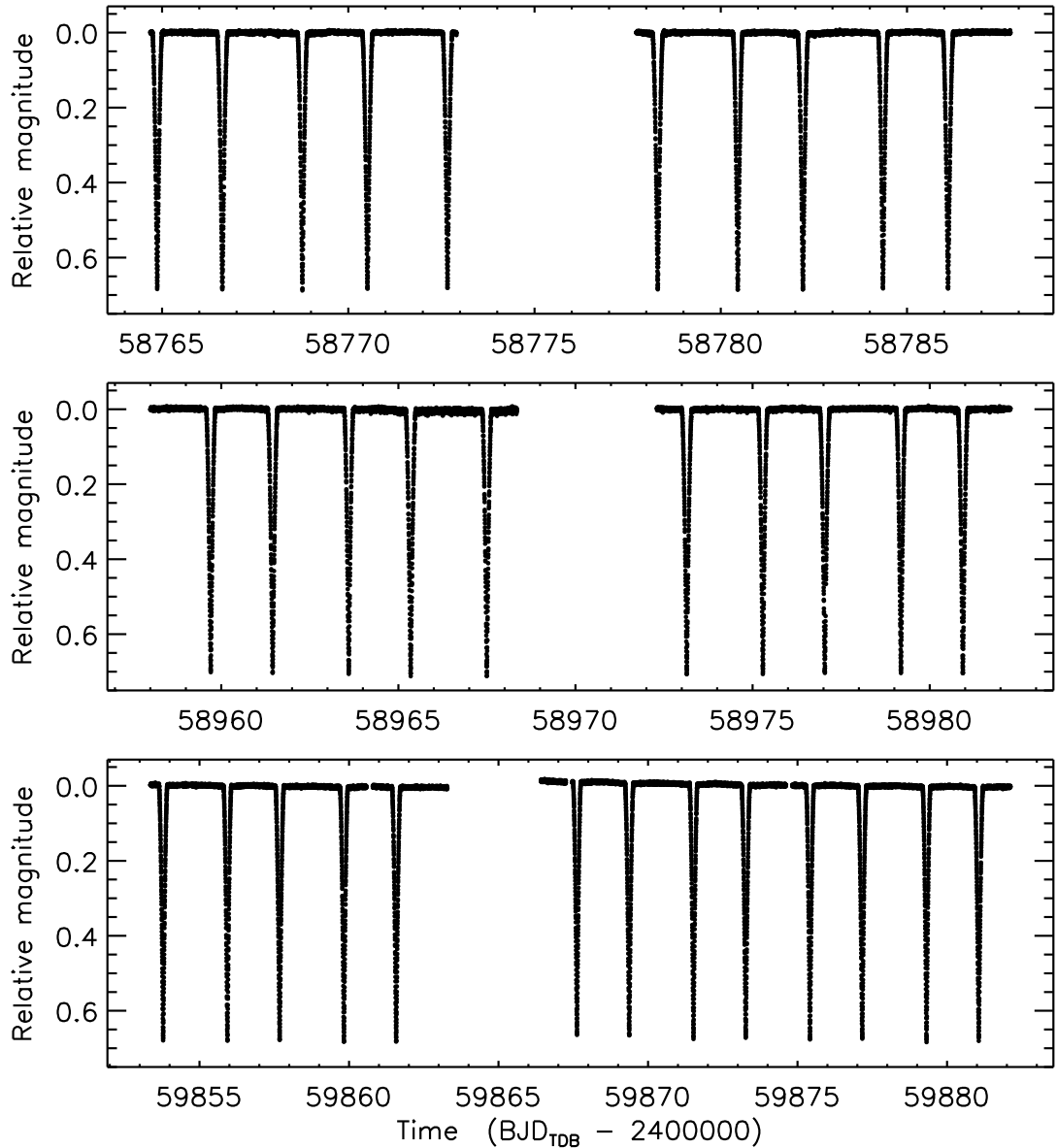} \\
\caption{\label{fig:time} TESS\ short-cadence SAP photometry of \targ\ from sectors 17 (top),
24 (middle) and 57 (bottom). The flux measurements have been converted to magnitude units
then rectified to zero magnitude by the subtraction of quadratic functions.} \end{figure}


The TESS mission\cite{Ricker+15jatis} observed \targ\ in sectors 17 (2019/10/07 to 2019/11/02), 24 (2020/04/16 to 2020/05/13) and 57 (2022/09/30 to 2022/10/29). All three sectors were observed in short cadence mode with a 120~s sampling rate. The simple aperture photometry (SAP) and pre-search data conditioning SAP (PDCSAP) data \cite{Jenkins+16spie} are almost indistinguishable, so we used the SAP data in our analysis for consistency with previous papers in this series. The eclipses in the PDC light curves are approximately 0.05~mag deeper than in the SAP light curves, indicating that the PDC data have been corrected for contaminating light. We prefer to model the SAP data and fit for third light, on the grounds that it is better to adjust the model to match the data than adjust the data to match the model.

The reduced data were downloaded from the MAST archive\footnote{Mikulski Archive for Space Telescopes, \\ \texttt{https://mast.stsci.edu/portal/Mashup/Clients/Mast/Portal.html}} and converted from flux units to relative magnitude. We required a QUALITY flag of zero, which yielded 12\,942 of the 18\,012 datapoints from sector 17, 16\,309 of 19\,074 from sector 24, and 17\,990 of 20\,712 from sector 57. We further trimmed the data from sector 24 to remove parts of the light curve associated with incompletely-observed eclipses, leaving 14\,459 datapoints for further analysis (Fig.~\ref{fig:time}). We did not use the errors provided with the datapoints as they are too small. We preferred instead to determine the precision of the photometry from the scatter around the best-fitting model.

\begin{figure}[t] \centering \includegraphics[width=\textwidth]{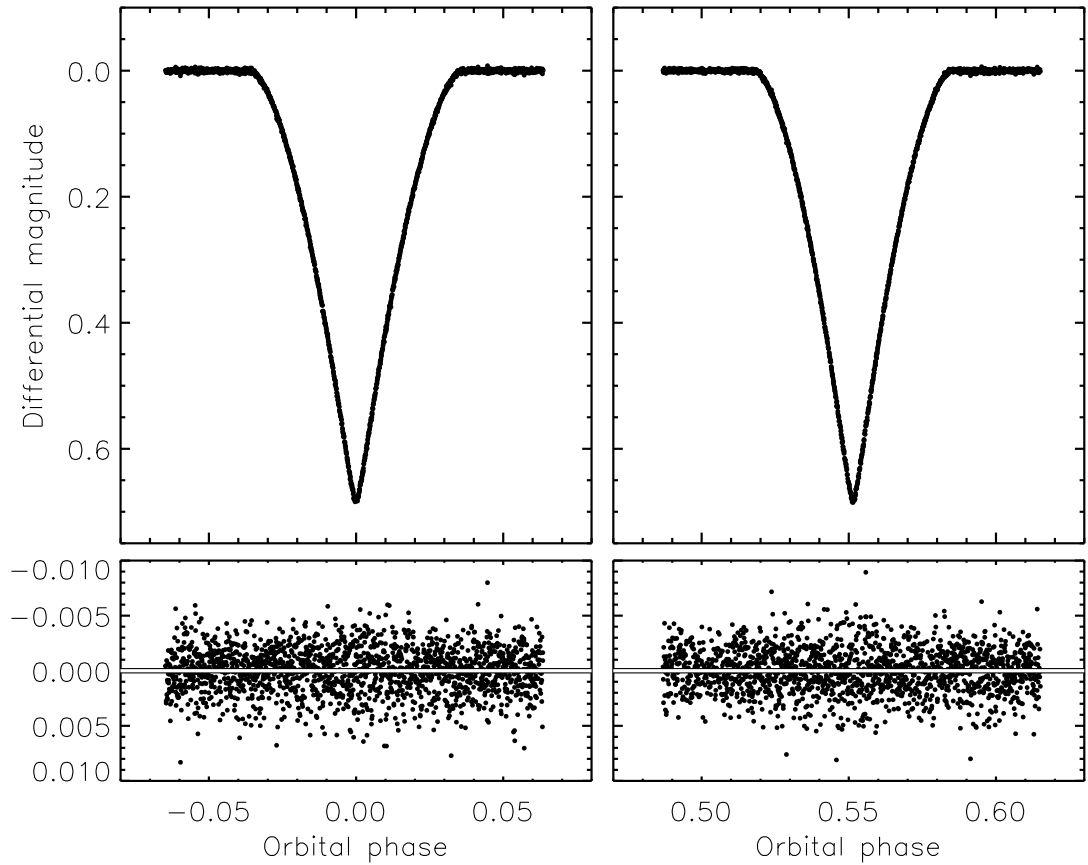} \\
\caption{\label{fig:phase17} Best fit to the TESS sector 17 light curve of \targ\
using {\sc jktebop} for the primary (left) and secondary (right) eclipses. The
residuals are shown on an enlarged scale in the lower panels.} \end{figure}

We queried the \gaia\ DR3 database\footnote{\texttt{https://vizier.cds.unistra.fr/viz-bin/VizieR-3?-source=I/355/gaiadr3}} in the region of \targ. A total of 146 additional sources are listed within 2~arcmin -- the constellation of Cassiopeiae is close to the galactic plane so has a relatively high surface density of point sources. The brightest of these is fainter than \targ\ by 4.31~mag in the \Grp\ passband (a light ratio of 1.9\%). This implies that there is a small but significant amount of contaminating light in the TESS data which must be accounted for in the light curve analysis.


\section*{Light curve analysis}

Inspection of the SAP light curves showed that the eclipse depths vary between the three TESS sectors. The primary eclipses are approximately 0.685~mag deep in sectors 17 and 57, and 0.705~mag deep in sector 24. The secondary eclipses show exactly the same behaviour. Apsidal motion is unlikely to cause this because the apsidal period is too long and the eclipse depths do not vary in antiphase. The probable explanation is that the different spacecraft orientations in the three sectors, combined with the relatively large 21\as\ pixel size, means that the amount of contaminating light changes. The three light curves should therefore be fitted separately.

\begin{figure}[t] \centering \includegraphics[width=\textwidth]{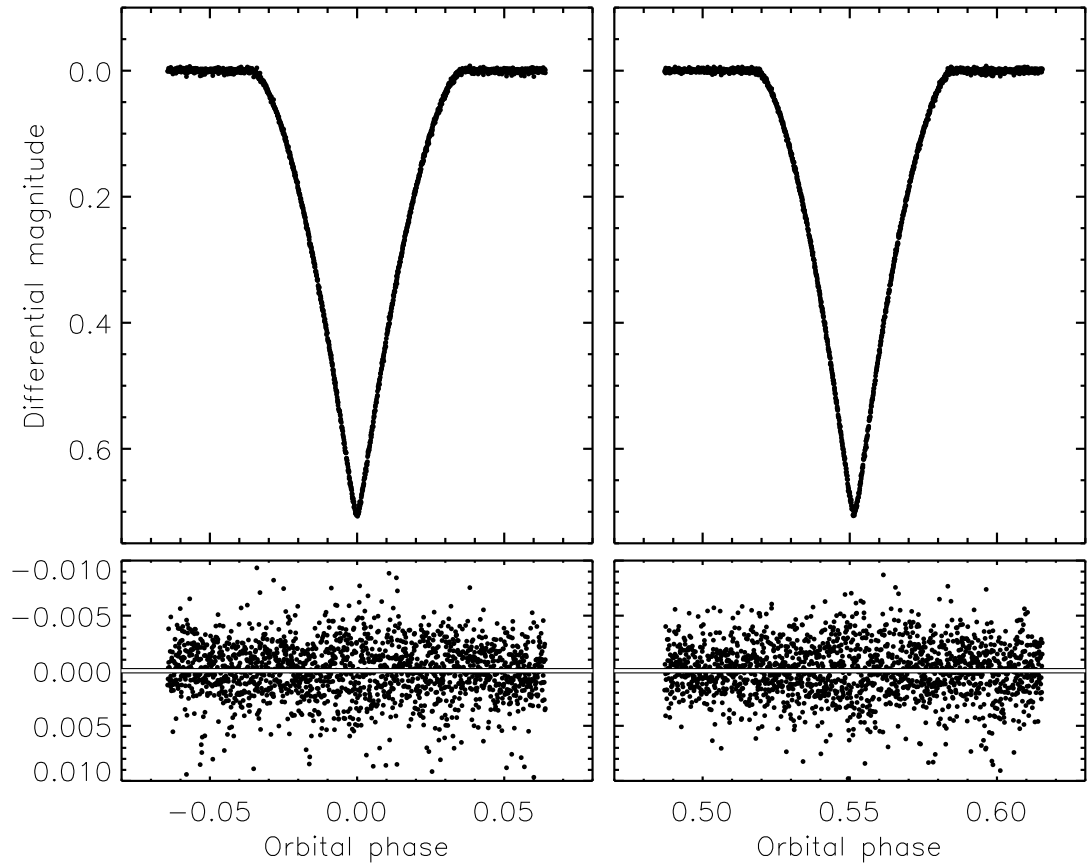} \\
\caption{\label{fig:phase24} Same as Fig.~\ref{fig:phase17}
but for the TESS data from sector 24.} \end{figure}

The light curves from the three sectors were each modelled using version 43 of the {\sc jktebop}\footnote{\texttt{http://www.astro.keele.ac.uk/jkt/codes/jktebop.html}} code \cite{Me++04mn2,Me13aa}. The parameters of the fit included the fractional radii of the stars, expressed as their sum ($r_{\rm A}+r_{\rm B}$) and ratio ($k = {r_{\rm B}}/{r_{\rm A}}$), the orbital inclination ($i$), the central surface brightness ratio ($J$), the amount of contaminating light ($L_3$) and the coefficients of the reflection effect. The orbital eccentricity ($e$) and argument of periastron ($\omega$) were included using the Poincar\'e elements ($e\cos\omega$ and $e\sin\omega$). The secondary eclipse was found to occur at orbital phase 0.552.

Following the results from Southworth \cite{Me23obs2} we included limb darkening using the power-2 law \cite{Hestroffer97aa} and the re-parameterisation into $h_1$ and $h_2$ given by Maxted \cite{Maxted18aa}. The two stars are almost identical so we forced them to have the same limb darkening coefficients, and fitted for both coefficients.

\targ\ exhibits slow apsidal motion, but a full analysis is beyond the scope of the current work. We therefore determined an orbital ephemeris separately for each TESS sector and did not interpret them further. The primary and secondary eclipses are of practically indistinguishable depth and we cannot confidently decide which is which. This contrasts with a similar situation we found for ZZ~Boo \cite{Me23obs1} where the TESS data definitively determined -- for the first time -- which of the two types of eclipses was deeper and thus by definition the primary. In the case of \targ\ the eclipse depths are even more similar and also will change over the apsidal period as $\omega$ cycles round. We therefore adopted the same convention for eclipse identifications as L97. We refer to the star eclipsed at phase 0.0 as star~A and to its companion as star~B.

\begin{figure}[t] \centering \includegraphics[width=\textwidth]{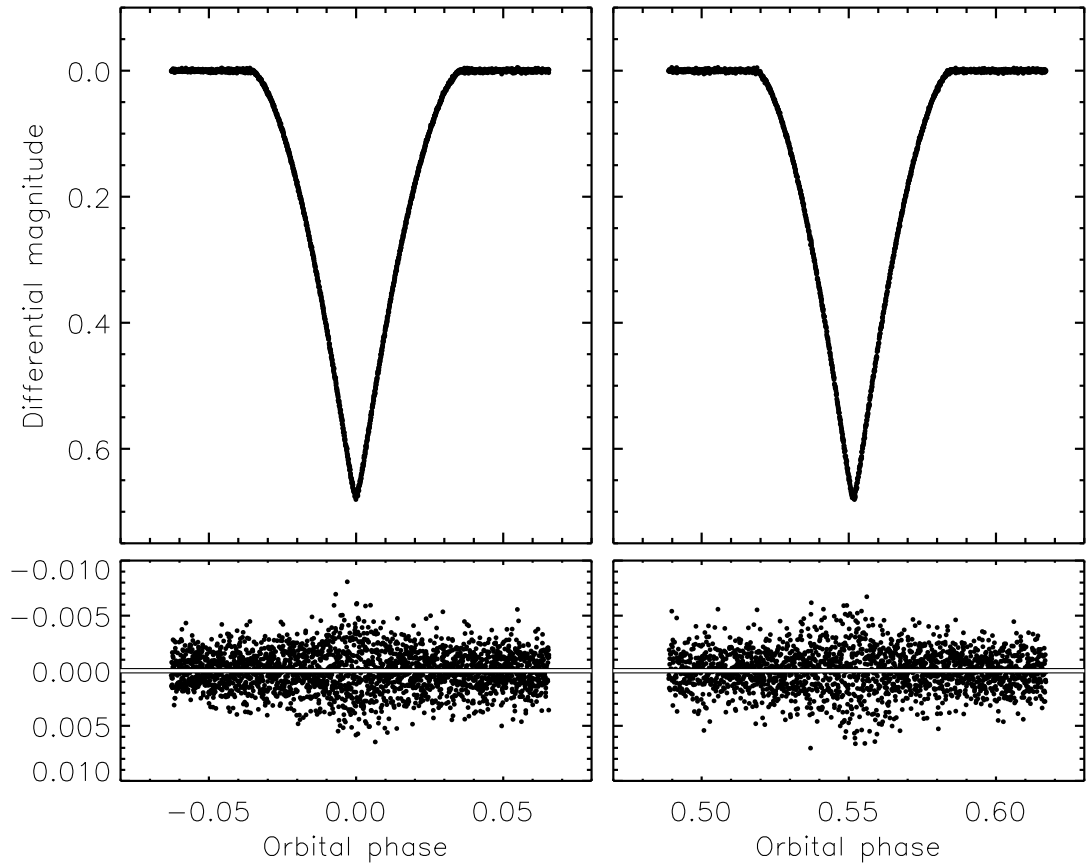} \\
\caption{\label{fig:phase57} Same as Fig.~\ref{fig:phase17}
but for the TESS data from sector 57.} \end{figure}

The best fits to the eclipse light curves are shown in Figs.\ \ref{fig:phase17}, \ref{fig:phase24} and \ref{fig:phase57}. Their parameters are given in Table~\ref{tab:jktebop}. Uncertainties in the parameters were determined using both  Monte Carlo and residual-permutation simulations as implemented in {\sc jktebop} \cite{Me++04mn,Me08mn}, the two alternatives being in close agreement for all parameters. The consistency between sectors is high, with the values for all but two parameters being in good agreement. Given this and the very small errorbars, we adopted the straight mean of the values for each parameter. We did the same for the errorbars, foregoing the division by $\sqrt{3}$ to convert to standard error. The values of $L_3$ are not in good agreement, as was expected given the change in eclipse depths between sectors. The disagreement in $e\cos\omega$ is stronger (reduced $\chi^2 = 49$) and remains unexplained: we have multiplied the final errorbar in this quantity by a factor of seven to account for the discrepancy.

\begin{sidewaystable} \centering
\caption{\em \label{tab:jktebop} Adopted parameters of \targ\ measured from the
TESS\ light curves using the {\sc jktebop} code. The uncertainties are 1$\sigma$
and were determined using Monte Carlo and residual-permutation simulations.}
\begin{tabular}{lcccc}
{\em Parameter}                           &       {\em Sector 17}              &       {\em Sector 24}              &       {\em Sector 57}              &         {\em Adopted}              \\[3pt]
{\it Fitted parameters:} \\
Time of primary eclipse (BJD$_{\rm TDB}$) & $   58778.308750 \pm  0.000013   $ & $   58973.141188 \pm  0.000017   $ & $   59869.370277 \pm  0.000015   $ &                                    \\
Orbital period (d)                        & $     3.896644   \pm  0.000005   $ & $     3.896638   \pm  0.000008   $ & $     3.896646   \pm  0.000005   $ &                                    \\
Orbital inclination (\degr)               & $       89.714   \pm  0.015      $ & $       89.679   \pm  0.022      $ & $       89.727   \pm  0.017      $ & $       89.707   \pm  0.018      $ \\
Sum of the fractional radii               & $       0.21568  \pm  0.00014    $ & $       0.21535  \pm  0.00023    $ & $       0.21558  \pm  0.00024    $ & $       0.21554  \pm  0.00020    $ \\
Ratio of the radii                        & $       0.9986   \pm  0.0021     $ & $       0.9944   \pm  0.0035     $ & $       0.9942   \pm  0.0050     $ & $       0.9957   \pm  0.0035     $ \\
$e\cos\omega$                             & $       0.080838 \pm  0.000007   $ & $       0.080877 \pm  0.000010   $ & $       0.081059 \pm  0.000033   $ & $       0.08092  \pm  0.00012    $ \\
$e\sin\omega$                             & $      -0.03644  \pm  0.00032    $ & $      -0.03662  \pm  0.00039    $ & $      -0.03624  \pm  0.00099    $ & $      -0.03643  \pm  0.00057    $ \\
Central surface brightness ratio          & $       0.99963  \pm  0.00054    $ & $       0.99995  \pm  0.00083    $ & $       1.00112  \pm  0.00086    $ & $       1.00023  \pm  0.00074    $ \\
Third light                               & $       0.0481   \pm  0.0030     $ & $       0.0222   \pm  0.0044     $ & $       0.0536   \pm  0.0096     $ &                                    \\
LD coefficient $c$                        & $       0.602    \pm  0.075      $ & $       0.540    \pm  0.038      $ & $       0.595    \pm  0.020      $ & $       0.579    \pm  0.044      $ \\
LD coefficient $\alpha$                   &              0.50 (fixed)          &              0.50 (fixed)          &              0.50 (fixed)          &                                    \\
{\it Derived parameters:} \\
Fractional radius of star~A               & $       0.10791  \pm  0.00013    $ & $       0.10798  \pm  0.00020    $ & $       0.10782  \pm  0.00030    $ & $       0.10790  \pm  0.00021    $ \\
Fractional radius of star~B               & $       0.10777  \pm  0.00014    $ & $       0.10737  \pm  0.00024    $ & $       0.10776  \pm  0.00029    $ & $       0.10763  \pm  0.00022    $ \\
Eccentricity                              & $       0.08867  \pm  0.00013    $ & $       0.08878  \pm  0.00015    $ & $       0.08879  \pm  0.00040    $ & $       0.08875  \pm  0.00023    $ \\
Argument of periastron ($^\circ$)         & $     335.74     \pm  0.19       $ & $     335.64     \pm  0.23       $ & $     335.91     \pm  0.60       $ & $     335.76     \pm  0.34       $ \\
Light ratio $\ell_{\rm B}/\ell_{\rm A}$   & $       0.9969   \pm  0.0038     $ & $       0.9887   \pm  0.0062     $ & $       1.0000   \pm  0.0085     $ & $       0.9952   \pm  0.0062     $ \\[3pt]
\end{tabular}
\end{sidewaystable}

\section*{Radial velocities}

\begin{figure}[t] \centering \includegraphics[width=\textwidth]{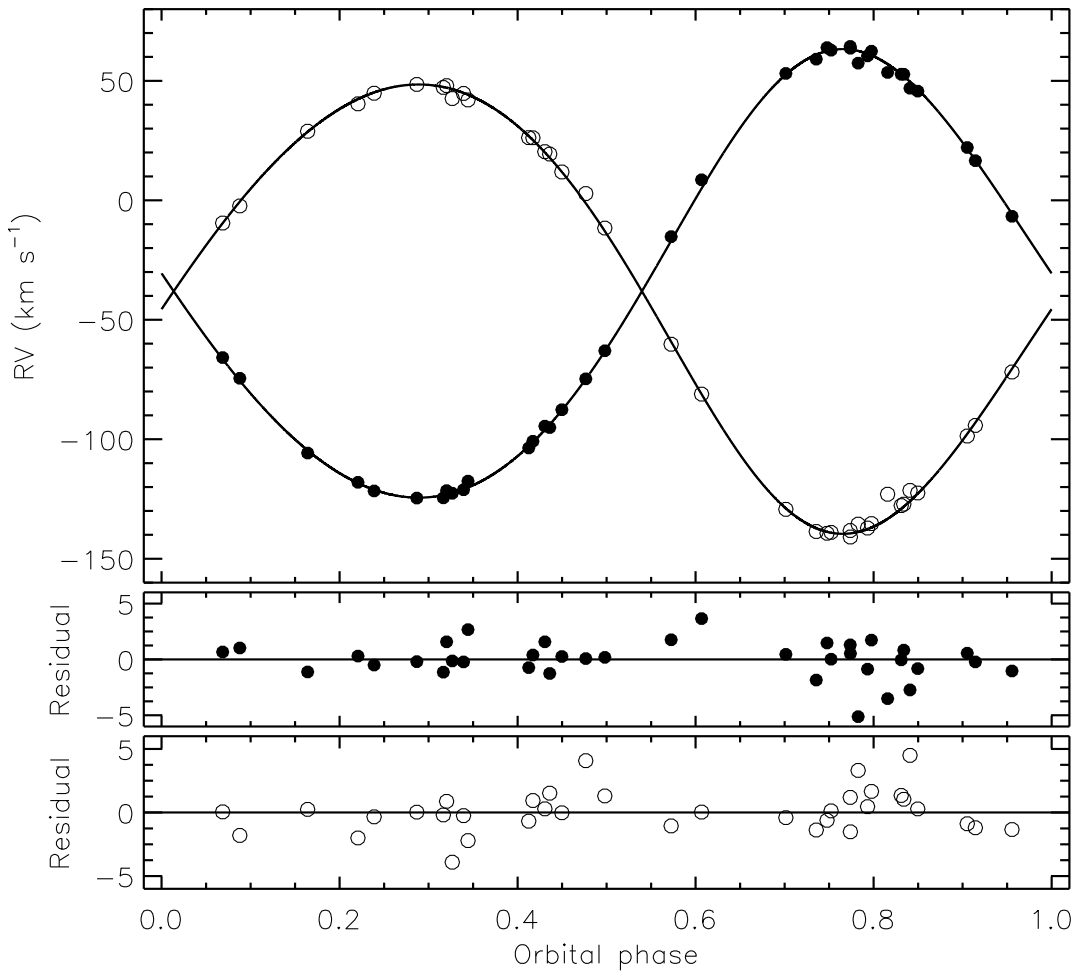} \\
\caption{\label{fig:rv} RVs of \targ\ from L97 (filled circles for star~A and
open circles for star~B) compared to the best-fitting spectroscopic orbits
from our own analysis using {\sc jktebop} (solid curves). The residuals are
given in the lower panels separately for the two components.} \end{figure}

We are aware of only one published spectroscopic study of IT~Cas: that of L97. We copied the RVs from that work and fitted them ourselves to confirm the results. Errorbars were assigned to the two different sources of RVs (``CHSL'' and ``CfA'') according to the standard errors given in table~10 of L97, and were subsequently adjusted by a small amount to force $\chir = 1.0$. A solution to the RVs was obtained with {\sc jktebop}, fitting for the velocity amplitudes ($K_{\rm A}$ and $K_{\rm B}$) and systemic velocities ($V_{\rm \gamma,A}$ and $V_{\rm \gamma,B}$) of the two stars, and the reference time of mid-eclipse. The quantities $e\cos\omega$ and $e\sin\omega$ were fixed to the values given in Table\,\ref{tab:jktebop} -- an alternative solution with these parameters fitted returned values of $K_{\rm A}$ and $K_{\rm B}$ larger by an insignificant 0.05\kms.

Uncertainties were determined using Monte Carlo simulations \cite{Me21obs5} and found to be slightly larger than the formal errors from the covariance matrix. Our results are: $K_{\rm A} = 93.85 \pm 0.24$\kms, $K_{\rm B} = 94.03 \pm 0.31$\kms, $V_{\rm \gamma,A} = -38.16 \pm 0.18$\kms\ and $V_{\rm \gamma,B} = -37.94 \pm 0.24$\kms. These agree very well with the value from L97, the main difference in the two analyses being our use of newer and more precise $e\cos\omega$ and $e\sin\omega$ values. A plot of the spectroscopic orbits is given in Fig.~\ref{fig:rv}.

%
%


\section*{Search for pulsations}

Previous studies of \targ\ have claimed the detection of $\delta$~Scuti pulsations in either star~A \cite{KhaliullinKozyreva89apss} or star~B \cite{HolmgrenWolf96obs}, but their presence was not confirmed (L97). Pulsations in dEBs are widespread \cite{Kahraman+17mn,GaulmeGuzik19aa,Shi+22apjs} and are important laboratories for stellar physics \cite{Bowman+19apj,Me21obs6,MeBowman22mn}. The components of \targ\ have \Teff s and masses within the lower half of the $\delta$~Scuti instability strip \cite{Uytterhoeven+11aa,Murphy+19mn}. We therefore searched for pulsations in the TESS light curve of this system.

This was done by performing a frequency analysis on the residuals of the best fit to the full light curves from the three TESS sectors individually, covering frequencies from 0 to 100~d$^{-1}$. The sectors were not combined as this would have led to strong aliasing effects. We find no evidence for pulsations within the frequency range considered, to a 3$\sigma$ upper limit of 0.10~mmag. Although $\delta$~Scuti stars do show variations in pulsation amplitude, the likely explanation is that the features seen in the older light curves of \targ\ are red noise rather than of astrophysical origin.

Kozyreva \etal\ \cite{KozyrevaZakharov01astl} found slow variations in the brightness of \targ\ on a monthly timescale. The TESS data are not well suited to the detection of periodicity on such long timescales, so we did not investigate this possibility further.


\section*{Physical properties of \targ}

\begin{table} \centering
\caption{\em Physical properties of \targ\ defined using the nominal solar units given by IAU
2015 Resolution B3 (ref.\ \cite{Prsa+16aj}). \label{tab:absdim}}
\begin{tabular}{lr@{\,$\pm$\,}lr@{\,$\pm$\,}l}
{\em Parameter}        & \multicolumn{2}{c}{\em Star A} & \multicolumn{2}{c}{\em Star B}    \\[3pt]
Mass ratio   $M_{\rm B}/M_{\rm A}$          & \multicolumn{4}{c}{$0.9981 \pm 0.0042$}       \\
Semimajor axis of relative orbit (\Rsunnom) & \multicolumn{4}{c}{$14.414 \pm 0.030$}        \\
Mass (\Msunnom)                             &  1.3244 & 0.0094      &  1.3218 & 0.0080      \\
Radius (\Rsunnom)                           &  1.5552 & 0.0044      &  1.5513 & 0.0045      \\
Surface gravity ($\log$[cgs])               &  4.1765 & 0.0022      &  4.1778 & 0.0021      \\
Density ($\!\!$\rhosun)                     &  0.3521 & 0.0022      &  0.3540 & 0.0023      \\
Synchronous rotational velocity ($\!\!$\kms)& 20.19   & 0.06        &  20.14  & 0.06        \\
Effective temperature (K)                   &   6740  & 105         &   6740  & 105         \\
Luminosity $\log(L/\Lsunnom)$               &   0.653 & 0.027       &   0.651 & 0.027       \\
$M_{\rm bol}$ (mag)                         &   3.108 & 0.068       &   3.113 & 0.068       \\
Distance (pc)                               & \multicolumn{4}{c}{$505.5 \pm 8.3$}           \\[3pt]
\end{tabular}
\end{table}

The physical properties of \targ\ were determined using the {\sc jktabsdim} code \cite{Me++05aa}, the measured values of $r_{\rm A}$, $r_{\rm B}$, $i$, $P$ and $e$ from Table~\ref{tab:jktebop}, and the $K_{\rm A}$ and $K_{\rm B}$ from above. The results are given in Table~\ref{tab:absdim}, where the errorbars have been propagated from all input parameters using a perturbation approach. The uncertainties on the radii of the stars are 0.3\%, which is slightly greater than the lower limit of 0.2\% to which results are expected to be reliable \cite{Maxted+20mn}. The largest source of uncertainty for both the masses \emph{and radii} of the stars is the uncertainty in the velocity amplitudes, thanks to the high quality of the TESS light curve. The two stars are identical to within the errorbars, with differences in mass of $0.3 \pm 1.2$\% and in radius of $0.5 \pm 0.6$\%. The agreement between our results and those of L97 is good but slightly less than expected given the errorbars.


L97 gave a \Teff\ value for both stars of $6740 \pm 105$~K, somewhat larger than the values of $6579 \pm 135$~K in the TESS Input Catalog (TICv8 \cite{Stassun+19aj}) and $6330 \pm 8$~K in \gaia\ DR3. To check this we determined the distance to the system using the {\sc jktabsdim} code, the $UBV$ magnitudes from Lacy \cite{Lacy92aj}, the $JHK_s$ magnitudes from 2MASS \cite{Cutri+03book} converted to the Johnson system using the transformations from Carpenter \cite{Carpenter01aj}, an interstellar reddening of $E(B-V) = 0.083 \pm 0.056$\,mag from the {\sc stilism}\footnote{\texttt{https://stilism.obspm.fr}} online tool \cite{Lallement+14aa,Lallement+18aa}, and the surface brightness versus \Teff\ relations from Kervella et al.\ \cite{Kervella+04aa}. This yielded $505.5 \pm 8.3$~pc, which compares well with the $515.0 \pm 4.4$~pc from simple inversion of the parallax of the system from \gaia\ DR3. The lower \Teff\ from TICv8 is ruled out to 2$\sigma$, and the \gaia\ DR3 \Teff\ to higher confidence. Based on this, we accept the \Teff\ from L97 as suitable for both stars.

The adopted \Teff\ value corresponds to a spectral class of F3 on the scale of Pecaut \& Mamajek \cite{PecautMamajek13apjs}. We therefore infer a spectral type of the system of F3\,V + F3\,V. This is somewhat earlier than the F5 mentioned by L97 and the F6 given by {\it Simbad}\footnote{\texttt{http://simbad.u-strasbg.fr/simbad/sim-id?Ident=it+cas\&submit=submit+id}} (without reference). This spectral type was arrived at from the \Teff\ and evolutionary stage of the stars, and is not a true spectral classification.


\section*{Summary and conclusions}

\targ\ is a dEB containing two F3~V stars on a 3.90~d orbit with a small orbital eccentricity. The two stars are identical in mass, radius and \Teff\ to within the uncertainties. TESS observed the system during three sectors covering approximately 3~yr, giving light curves of very high quality. We have modelled these data using the {\sc jktebop} code to determine the photometric properties of the system. We also analysed the RVs of \targ\ published by L97, finding results in good agreement with that work. From the measured parameters we have calculated the physical properties of the system (Table~\ref{tab:jktebop}) to precisions of 0.6\% in mass and 0.3\% in radius. The \Teff s found by L97 yield a distance measurement in full agreement with the parallax from \gaia\ DR3. We searched for and found no evidence for pulsations in the light curve.

As a sanity check we have compared the masses, radii and \Teff s of the stars to predictions from the {\sc parsec} stellar evolutionary models \cite{Bressan+12mn}. A fractional metal abundance by mass of $Z = 0.014$ and an age of $2.0 \pm 0.1$~Gyr provide a good match to the measured properties. \targ\ is now a well-understood dEB, but would benefit from high-resolution spectroscopy for the measurement of its photospheric chemical composition and more precise \Teff\ values. The system shows apsidal motion, and the measured apsidal period would be a useful addition to a detailed comparison between the properties of \targ\ and theoretical predictions.


\section*{Acknowledgements}

We are grateful to Steve Overall and an anonymous referee for useful comments on a draft of this work. This paper includes data collected by the TESS\ mission and obtained from the MAST data archive at the Space Telescope Science Institute (STScI). Funding for the TESS\ mission is provided by the NASA's Science Mission Directorate. STScI is operated by the Association of Universities for Research in Astronomy, Inc., under NASA contract NAS 5–26555.
 The following resources were used in the course of this work: the NASA Astrophysics Data System; the SIMBAD database operated at CDS, Strasbourg, France; and the ar$\chi$iv scientific paper preprint service operated by Cornell University.



\end{document}